\documentclass[twoside]{article}
\usepackage{amsfonts, amsthm, amssymb}
\usepackage[mathcal]{eucal}

\newtheorem{theorem}{Theorem}[section]
\newtheorem{lemma}[theorem]{Lemma}
\newtheorem{prop}[theorem]{Proposition}
\newtheorem{cor}[theorem]{Corollary}
\newtheorem{remark}[theorem]{Remark}

\def\l{\lambda}
\def\a{\alpha}
\def\b{\beta}

\def\g{\gamma}
\def\d{\delta}
\def\e{\varepsilon}

\def\z{\zeta}

\def\R{\mathbb{R}}

\def\C{\mathbb{C}}

\def\Z{\mathbb{Z}}
\def\P{\mathbb{P}}

\def\M{\mathcal{M}}
\def\Mba{\M_{\beta, \alpha}}
\def\Hba{H_{\beta, \alpha}}
\def\Rba{R_{\beta, \alpha}}

\def\P{\mathcal{P}}
\def\Pba{\P_{\beta, \alpha}}

\def\Oba{\Omega_{\beta, \alpha}}

\def\hrba{\Hba \circ \Rba}
\def\hcalba{\mathcal{H}_{\b,\a}}
\def\cbao{C_{\b,\a}^\omega}

\def\n{\nabla}
\def\nba{\nabla_{b,a}}

\def\mbf2{\mathbf{2}}

\def\1N1{1 \leq k \leq N-1}

\def\idplgeq{``Id $+$ higher order terms''}

\begin{document}

\title{Birkhoff normal form for the periodic Toda lattice}
\author{Andreas Henrici\footnote{Supported in part by the Swiss National Science Foundation} \and Thomas Kappeler\footnote{Supported in part by the Swiss National Science Foundation, the programme SPECT and the European Community through the FP6 Marie Curie RTN ENIGMA (MRTN-CT-2004-5652)}}

\maketitle

\begin{center}
\emph{To Percy Deift at the occasion of his $60$'th birthday}
\end{center}

\begin{abstract}
In this paper we compute the Birkhoff normal form of the periodic Toda lattice up to order four. As an application, we verify that Kolmogorov's nondegeneracy condition in the KAM theorem holds almost everywhere in phase space.
\end{abstract}

\section{Introduction}

Consider the periodic Toda lattice with period $N$ ($N \geq 2$), 
\begin{displaymath}
\dot{q}_n = \partial_{p_n} H, \quad \dot{p}_n = -\partial_{q_n} H
\end{displaymath}
for $n \in \Z$, where the (real) coordinates $(q_n, p_n)_{n \in \Z}$ satisfy $(q_{n+N}, p_{n+N}) = (q_n, p_n)$ for any $n \in \Z$ and the Hamiltonian $H$ is given by
\begin{displaymath}
  H = \frac{1}{2} \sum_{n=1}^N p_n^2 + \sum_{n=1}^N V(q_n - q_{n+1})
\end{displaymath}
with potential $V(x) = \g^2 e^{\d x} + V_1 x + V_2$ and $\g$, $\d$, $V_1$, $V_2 \in \R$ constants. The Toda lattice has been introduced by Toda \cite{toda} and studied extensively in the sequel. It is an FPU lattice, i.e. a Hamiltonian system of particles in one space dimension with nearest neighbor interaction. Models of this type have been studied by Fermi-Pasta-Ulam [FPU]. In numerical experiments they found recurrent features for the lattices they considered. Despite an enormous effort from the physics and mathematics community, some of these numerical experiments still defy an explanation. For a recent account of the fascinating history of the FPU problem, see e.g. \cite{beiz}. At least in the case of the periodic Toda lattice, the recurrent features can be fully accounted for. In fact, Flaschka \cite{fla1}, H\'enon \cite{henon}, and Manakov \cite{mana} independently proved that the periodic Toda lattice is integrable. In this paper, we show that (small) Hamiltonian perturbations of the Toda lattice have many - large and small - quasi-periodic solutions as well. To this end, we put the periodic Toda lattice into Birkhoff normal form up to order $4$ and then show that Kolmogorov's nondegeneracy condition of the KAM theorem holds almost everywhere in phase space - see e.g. \cite{kapo}, Appendix G, for the notion of Birkhoff normal form up to order $m$. 

Returning to the Toda Hamiltonian $H$, let us first note that w.l.o.g. we can assume that $V_1 = V_2 = 0$. When expressed in the canonical coordinates $(\d q_j, \frac{1}{\d} p_j)_{1 \leq j \leq N}$ the Hamiltonian $H$ is, up to a scaling factor $\d^{-2}$, of the form
\begin{equation} \label{htoda}
H_{Toda} = \frac{1}{2} \sum_{n=1}^N p_n^2 + \alpha^2 \sum_{n=1}^N e^{q_n - q_{n+1}},
\end{equation}
where $\a^2 = (\g \d)^2$. Moreover notice that the total momentum $\sum_{n=1}^N p_n$ is conserved; hence the motion of the center of mass $\frac{1}{N} \sum_{n=1}^N q_n$ is linear and therefore unbounded. However, it turns out that the orbits of the system relative to the center of mass all lie on tori, generically of dimension $N-1$. To describe these orbits, it suffices to consider the relative coordinates $v_n := q_{n+1} - q_n$ ($1 \leq n \leq N-1$) and their conjugate ones, $u_n := n\b - \sum_{j=1}^n p_k$ ($1 \leq n \leq N-1$), where $\b = \frac{1}{N} \sum_{j=1}^N p_n$. The corresponding phase space is then $\R^{2N-2}$.

In terms of the variables $(v_k, u_k)_{\1N1}$, $H_{Toda}$ is given by
\begin{equation} \label{hba(vu)}
  \Hba \!=\! \frac{N\b^2}{2} \!+\! \frac{1}{2} \left( \! u_1^2 \!+\! \sum_{l=1}^{N-2} (u_l \!-\! u_{l\!+\!1})^2 \!+\! u_{N-1}^2 \! \right) \!+\! \a^2 \left( \! \sum_{k=1}^{N-1} e^{-v_k} + e^{\sum_{k=1}^{N-1} v_k} \! \right),
\end{equation}
with parameters $\a$ and $\b$ - see section \ref{relcoord} for more details.

The main result of this paper is the following
\begin{theorem} \label{bnftodatheorem}
For any fixed $\b \in \R$, $\a > 0$, and $N \geq 2$, the periodic Toda lattice admits a Birkhoff normal form. More precisely, there are (globally defined) canonical coordinates $(x_k, y_k)_{\1N1}$ so that $\Hba$, when expressed in these coordinates, takes the form $\hcalba(I) := \frac{N\b^2}{2} + H_\a(I)$, where $H_\a(I)$ is a real analytic function of the action variables $I_k = (x_k^2 + y_k^2)/2$ ($\1N1$). Moreover, near $I=0$, $H_\a(I)$ has an expansion of the form
\begin{equation} \label{halphataylor}
  N \a^2 + \a \sum_{k=1}^{N-1} s_k I_k + \frac{1}{4N} \sum_{k=1}^{N-1} I_k^2 + O(I^3),
\end{equation}
with $s_k = 2 \sin \frac{k \pi}{N}$ for $\1N1$.
\end{theorem}

\begin{remark}
In particular, Theorem \ref{bnftodatheorem} says that near $0$, the Toda lattice can be viewed as a system of $(N-1)$ nonlinear harmonic oscillators which are uncoupled up to order $4$. The system is resonant at $0$ with resonance lattice $\mathcal{R}es := \{ (k_j)_{1 \leq j \leq N-1} \in \Z^{N-1} | \sum_{j=1}^{N-1} k_j s_j = 0 \}$ of dimension at least $\llcorner \frac{N-1}{2} \lrcorner$ as $s_k - s_{N-k} = 0$ for any $1 \leq k \leq \llcorner \frac{N-1}{2} \lrcorner$.
\end{remark}

\begin{cor} \label{todahessian}
Let $\a > 0$ and $\b \in \R$ be arbitrary. Then the Hessian of $\hcalba(I)$ at $I=0$ is given by
\begin{displaymath}
  d^2_I \hcalba|_{I=0} = \frac{1}{2N} \textrm{Id}_{N-1}.
\end{displaymath}
In particular, the frequency map $I \mapsto \n_I \hcalba$ is nondegenerate at $I=0$ and hence, by analyticity, nondegenerate on an open dense subset of $(\R_{\geq 0})^{N-1}$.
\end{cor}

Theorem \ref{bnftodatheorem} and Corollary \ref{todahessian} allow to apply the KAM-theorem (cf e.g. \cite{kapo, poeschelkam82}) to the Toda lattice on an open dense subset of the phase space. Moreover, as the Hessian of $\hcalba$ at $I=0$ is positive definite and hence $\hcalba$ is strictly convex near $I=0$, one can also apply Nekhoroshev's theorem (cf e.g. \cite{poeschelnekh2}) to the Toda lattice near the equilibirum point.

Using the method of proof of this paper we computed in \cite{ahtk5} the Birkhoff normal form up to order $4$ for any FPU chain and applied these computations to improve on results of Rink \cite{rink01} with regard to the verification of Kolmogorov's condition near the equilibrium point for these systems.

\emph{Acknowledgement:} It is a great pleasure to thank Percy Deift and Yves Colin de Verdi\`ere for helpful discussions.

\section{Relative coordinates} \label{relcoord}

To prove the integrability of the Toda lattice, Flaschka introduced in \cite{fla1} the (noncanonical) coordinates
\begin{equation} \label{flaschkadef}
b_n := -p_n \in \R, \quad a_n := \alpha e^{\frac{1}{2} (q_n - q_{n+1})} \in \R_{>0} \quad (n \in \Z).
\end{equation}
They are coordinates which describe the motion of the Toda lattice relative to the center of mass. In these coordinates, the Hamiltonian $H_{Toda}$ takes the simple form $H = \frac{1}{2} \sum_{n=1}^N b_n^2 + \sum_{n=1}^N a_n^2$, and the equations of motion are
\begin{equation} \label{flaeqn}
\left\{ \begin{array}{lllll}
 \dot{b}_n & = & a_n^2 - a_{n-1}^2 \\
 \dot{a}_n & = & \frac{1}{2} a_n (b_{n+1} - b_n)
\end{array} \right. \quad (n \in \Z).
\end{equation}
Note that $(b_{n+N}, a_{n+N}) = (b_n, a_n)$ for any $n \in \Z$, and $\prod_{n=1}^N a_n = \alpha^N$. Hence we can identify the sequences $(b_n)_{n \in \Z}$ and $(a_n)_{n \in \Z}$ with the vectors $(b_n)_{1 \leq n \leq N} \in \R^N$ and $(a_n)_{1 \leq n \leq N} \in \R_{>0}^N$. The phase space of the system (\ref{flaeqn}) is then given by
\begin{displaymath}
\M := \R^N \times \R_{>0}^N,
\end{displaymath}
and it turns out that (\ref{flaeqn}) is a Hamiltonian system with respect to the nonstandard Poisson structure $J \equiv J_{b,a}$, defined at a point $(b,a) = (b_n, a_n)_{1 \leq n \leq N}$ by
\begin{displaymath}
J = \left( \begin{array}{cc}
0 & A \\
-A^T & 0 \\
\end{array} \right)
\end{displaymath}
where $A$ is the $b$-independent $N \times N$-matrix
\begin{equation} \label{adef}
A = \frac{1}{2} \left( \begin{array}{ccccc}
a_1 & 0 & \ldots & 0 & -a_N \\
-a_1 & a_2 & 0 & \ddots & 0 \\
0 & -a_2 & a_3 & \ddots & \vdots \\
\vdots & \ddots & \ddots & \ddots & 0 \\
0 & \ldots & 0 & -a_{N-1} & a_N \\
\end{array} \right).
\end{equation}
The Poisson bracket corresponding to $J$ is then given by
\begin{eqnarray}
\{ F, G \}_J(b,a) & = & \langle (\nabla_b F, \nabla_a F), J \, (\nabla_b G, \nabla_a G) \rangle_{\R^{2N}} \nonumber\\
& = & \langle \nabla_b F, A \, \nabla_a G \rangle_{\R^N} - \langle \nabla_a F, A^T \, \nabla_b G \rangle_{\R^N} \label{poisson}
\end{eqnarray}
where $F,G \in C^1(\M)$ and where $\nabla_b$ and $\nabla_a$ denote the gradients with respect to $b = (b_1, \ldots, b_N)$ and $a = (a_1, \ldots, a_N)$, respectively. Therefore, equations (\ref{flaeqn}) can alternatively be written as $\dot{b}_n = \{ b_n, H \}_J$, $\dot{a}_n = \{ a_n, H \}_J$ $(1 \leq n \leq N)$.

Since the matrix $A$, defined by (\ref{adef}), has rank $N-1$, the Poisson structure $J$ is degenerate. It admits the two Casimir functions\footnote{A smooth function $C: \M \to \R$ is a Casimir function for $J$ if $\{ C, \cdot \}_J \equiv 0$.}
\begin{equation} \label{casimirdef}
C_1 := -\frac{1}{N} \sum_{n=1}^N b_n \quad \textrm{and} \quad C _2 := \left( \prod_{n=1}^N a_n \right)^\frac{1}{N}.
\end{equation}
The gradients $\nba C_i \! = \! (\n_b C_i, \n_a C_i)$ $(i = 1,2)$ are given by
\setlength\arraycolsep{1.5pt} {
\begin{eqnarray}
\nabla_b C_1 & = & -\frac{1}{N} (1, \ldots, 1), \qquad \nabla_a C_1 = 0, \label{c1grad} \\
\nabla_b C_2 & = & 0, \qquad \nabla_a C_2 = \frac{C_2}{N} \left( \frac{1}{a_1}, \ldots, \frac{1}{a_N} \right). \label{c2grad}
\end{eqnarray}}
They are linearly independent at each point $(b,a) \in \M$.

Let $\Mba := \{ (b,a) \in \M : (C_1, C_2) = (\b, \a) \}$ denote the level set of $(C_1, C_2)$ at $(\b, \a) \in \R \times \R_{>0}$. As $C_1$ and $C_2$ are real analytic on $\M$ and the gradients $\nba C_1$ and $\nba C_2$ are linearly independent everywhere on $\M$, the sets $\Mba$ are real analytic submanifolds of $\M$ of (real) codimension two. Furthermore the Poisson structure $J$, restricted to $\Mba$, becomes nondegenerate everywhere on $\Mba$ and therefore induces a symplectic structure on $\Mba$. In this way, we obtain a symplectic foliation of $\M$ with $\Mba$ being the symplectic leaves. We denote by $\Hba$ the restriction of the Hamiltonian $H_{Toda}$ to $\Mba$.

Besides Flaschka's coordinates of the Toda lattice we will also need to consider relative coordinates. Introduce $(v_1, \ldots, v_N) \in \R^N$ given by $v_i := q_{i+1} - q_i$ for $1 \leq i \leq N-1$ and $v_N := \frac{1}{N} \sum_{i=1}^N q_i$. Then $v = M q$ is a linear change of the position coordinates where $M$ is the $N \times N$-matrix
\begin{displaymath}
M = \left( \begin{array}{ccccc}
-1 & 1 & 0 & \ldots & 0 \\
0 & \ddots & \ddots &  & \vdots \\
\vdots &&&& 0 \\
0 & & \ldots & -1 & 1 \\
N^{-1} & \ldots &  & \ldots & N^{-1} \\
\end{array} \right).
\end{displaymath}

The variables $u = (u_1, \ldots, u_N)$ conjugate to $v = (v_1, \ldots, v_N)$ are then given by $u = (M^T)^{-1} p$. The matrix $(M^T)^{-1}$ can be computed to be
\begin{displaymath}
(M^T)^{-1} = \frac{1}{N} \left( \begin{array}{ccccc}
1 & \ldots & \quad & \ldots & 1 \\
2 & \ldots &  & \ldots & 2 \\
\vdots & & & & \vdots \\
\vdots & & & & \vdots \\
N & \ldots &  & \ldots & N \\
\end{array} \right) - \left( \begin{array}{ccccc}
1 & 0 & \ldots & \ldots & 0 \\
1 & 1 & 0 & \ldots & 0 \\
\vdots & & & & \vdots \\
1 & \ldots & & 1 & 0 \\
0 & \ldots &  & \ldots & 0 \\
\end{array} \right).
\end{displaymath}

For any $(\b, \a) \in \R \times \R_{>0}$ the variables $(v_k, u_k)_{1 \leq k \leq N-1}$ are canonically conjugate variables on $\Mba$. We want to express the Hamiltonian $\Hba$ in terms of the coordinates $(v,u) = (v_k, u_k)_{1 \leq k \leq N-1}$. Note that on $\Mba$, $u_k = k\b - \sum_{j=1}^k p_j$ for $\1N1$ and $u_N = N\b$.
Hence $p_1 = -u_1 + \b$, $p_N = u_{N-1} + \b$, and $p_k = (u_{k-1} - u_k) + \b$ for $2 \leq k \leq N-1$,
and thus
\begin{displaymath}
\frac{1}{2} \sum_{j=1}^N p_j^2 = \frac{N \b^2}{2} + \frac{1}{2} \left( u_1^2 + (u_1 - u_2)^2 + \ldots + (u_{N-2} - u_{N-1})^2 + u_{N-1}^2 \right).
\end{displaymath}
Moreover, using that $q_N - q_{N+1} = q_N - q_1 = \sum_{k=1}^{N-1}(q_{k+1} - q_k)$ one gets
\begin{displaymath}
  \sum_{j=1}^N e^{q_j - q_{j+1}} = \sum_{k=1}^{N-1} e^{-v_k} + e^{\sum_{k=1}^{N-1} v_k}.
\end{displaymath}
Combining the two expressions displayed above yields
\begin{equation} \label{hbavu}
 \!\! \Hba \!=\! \frac{N\b^2}{2} \!+\! \frac{1}{2} \left( \! u_1^2 \!+\! \sum_{l=1}^{N-2} (u_l \!-\! u_{l\!+\!1})^2 \!+\! u_{N-1}^2 \! \right) \!+\! \a^2 \left( \! \sum_{k=1}^{N-1} e^{-v_k} + e^{\sum_{k=1}^{N-1} v_k} \! \right).
\end{equation}

The coordinates $(v_k, u_k)_{1 \leq k \leq N-1}$ (as well as $u_N$, but \emph{not} $v_N$) can easily be expressed in terms of $(b_j, a_j)_{1 \leq j \leq N}$,
\begin{equation} \label{vuba}
  v_k = 2 \log \frac{\a}{a_k}, \quad u_k = -\frac{k}{N} \sum_{j=1}^N b_j + \sum_{j=1}^k b_j \qquad (1 \leq k \leq N-1)
\end{equation}
Note that $u_k$ depends linearly on the $b$-coordinates and is independent of the $a$-coordinates. On the other hand, $v_k$ depends only on the $a$-coordinates. The partial derivatives of the $v_k$'s at $a = \a 1_N$, where $1_N = (1, \ldots, 1) \in \R^N$, can be computed to be
\begin{equation} \label{dvkaj}
  \frac{\partial v_k}{\partial a_j} = -\frac{2}{\a} \delta_{kj} + \frac{2}{N\a}.
\end{equation}

\section{Birkhoff coordinates}

Let us recall from \cite{ahtk1} and \cite{ahtk2} our results concerning global Birkhoff coordinates on the phase space $\M$. As a model space, we introduced the space $\P := \R^{2N-2} \times \R \times \R_{>0}$, foliated by $\Pba := \R^{2N-2} \times \{ \beta \} \times \{ \a \}$ which are endowed with the standard symplectic structure. Denote by $J_0$ the degenerate Poisson structure on $\P$ having $\Pba$ as its symplectic leaves and the coordinates $\b$ and $\a$ as its Casimirs.

In \cite{ahtk1} we have proved that the Toda lattice admits global Birkhoff coordinates.
\begin{theorem} \label{sumthm}
There exists a map
\begin{displaymath}
\begin{array}{ccll}
 \Phi: & (\M, J) & \to & (\P, J_0) \\
 & (b,a) & \mapsto & ((x_k, y_k)_{1 \leq k \leq N-1}, C_1, C_2)
\end{array}
\end{displaymath}
with the following properties:
\begin{itemize}
  \item $\Phi$ is a real analytic diffeomorphism.
  \item $\Phi$ is canonical, i.e. it preserves the  Poisson brackets. In particular, the symplectic foliation of $\M$ by $\Mba$ is trivial.
  \item The coordinates $(x_k, y_k)_{1 \leq k \leq N-1}, C_1, C_2$ are
  global Birkhoff coordinates for the periodic Toda lattice, i.e. the transformed Toda Hamiltonian $H^* = H \circ \Phi^{-1}$
  is a function of the actions $(I_k)_{1 \leq k \leq N-1}$ and $C_1, C_2$ alone, where $I_k = \frac{1}{2} (x_k^2 + y_k^2)$ for any $ 1 \leq k \leq N-1$. (In the sequel we will denote $H^*$ by $H$ as well.)
\end{itemize}
\end{theorem}

As an immediate consequence of Theorem \ref{sumthm} one gets

\begin{cor} \label{respfol}
For every $\b \in \R$ and $\a > 0$,
\begin{displaymath}
\Phi(\Mba) = \Pba,
\end{displaymath}
and $\Phi|_{\Mba}: \Mba \to \Pba$ is a symplectomorphism. In particular, the moment map
\begin{displaymath}
  \Mba \to \R_{\geq 0}^{N-1}, \quad (b,a) \mapsto \big( I_k(\Phi(b,a)) \big)_{\1N1}
\end{displaymath}
is onto.
\end{cor}

We wish to analyze the Birkhoff map $\Phi$ and its restriction to the leaves $\Mba$ near the equilibrium points $(b,a) = (\b 1_N, \a 1_N)$, where $1_N = (1, \ldots, 1) \in \R^N$. Introduce for $k \in \Z$ with $1 \leq |k| \leq N-1$ the abbreviation
\begin{displaymath}
\l_k := \big| \sin \frac{k \pi}{N} \big|^\frac{1}{2}.
\end{displaymath}

To compute the Jacobian of $\Phi^{-1}$ at the points $(0_{N-1}, 0_{N-1},\b,\a)$ we first compute the Jacobian of $\Phi$. It turns out that the formulas are easier to express in complex notation. In \cite{ahtk2} we have shown the following
\begin{lemma} \label{birkdiff}
For any $\b \in \R$, $\a > 0$, and $\1N1$, the gradient $\nba (x_k + i y_k)$ at $(b, a) = (\b 1_N, \a 1_N)$ is given by
\begin{equation} \label{gradxnyn}
\nba (x_k + i y_k)(j) = \frac{1}{\sqrt{2 \a N}} \cdot \frac{1}{\l_k} \left\{  \begin{array}{ll}
e^{2\pi i(j-1)k/N} & (1 \leq j \leq N) \\
-2 e^{i\pi (2j-1)k/N} & (N+1 \leq j \leq 2N)
\end{array} \right.
\end{equation}
and
\begin{equation} \label{casgrad}
\nba C_1 = -\frac{1}{N}(1_N, 0_N), \quad \nba C_2 = \frac{1}{N}(0_N, 1_N).
\end{equation}
\end{lemma}

According to the formulas (\ref{gradxnyn}) and (\ref{casgrad}), the Jacobian $d_{b,a} \Phi$ of $\Phi$ at $(b,a) = (\b 1_N, \a 1_N)$ is given by the $(2N \times 2N)$-matrix
\begin{equation} \label{dbaxyc1c2}
\left( \begin{array}{c|c}
 & \\
\left( \frac{1}{\sqrt{2 \a N}} \frac{1}{\l_k} \cos\frac{(2j-2) \pi k}{N} \right)_{kj}  & \left( -\frac{2}{\sqrt{2 \a N}} \frac{1}{\l_k} \cos\frac{(2j-1) \pi k}{N} \right)_{kj} \\
&   \\
\hline
  & \\
\left( \frac{1}{\sqrt{2 \a N}} \frac{1}{\l_k} \sin\frac{(2j-2) \pi k}{N} \right)_{kj} & \left( -\frac{2}{\sqrt{2 \a N}} \frac{1}{\l_k} \sin\frac{(2j-1) \pi k}{N} \right)_{kj} \\
&  \\
\hline
-N^{-1} \ldots \qquad \qquad \ldots -N^{-1} & 0 \; \ldots \qquad \qquad \qquad \ldots \; 0 \\
\hline
0 \; \ldots \qquad \qquad \qquad \ldots \; 0  & N^{-1} \ldots \qquad \qquad \ldots  N^{-1}
\end{array} \right),
\end{equation}
where in each of the four $(N-1) \times N$-submatrices the row and column indices $k$ and $j$ run from $1$ to $N-1$ and $1$ to $N$, respectively. In order to compute its inverse, we note that (\ref{dbaxyc1c2}) can be written as the product $\Delta_2 \cdot P \cdot \Delta_1$ of the diagonal matrices
\begin{displaymath}
  \Delta_1 := \textrm{diag} \; (1_N, -2 \cdot 1_N)
\end{displaymath}
and
\begin{displaymath}
  \Delta_2 := \frac{1}{\sqrt{2\a}} \textrm{diag} \left( \frac{1}{\l_1}, \ldots, \frac{1}{\l_{N\!-\!1}}, \frac{1}{\l_1}, \ldots, \frac{1}{\l_{N\!-\!1}}, -\sqrt{\frac{2\a}{N}}, -\sqrt{\frac{\a}{2N}} \right)
\end{displaymath}
with the orthogonal $2N \times 2N$ matrix
\begin{displaymath}
P := \frac{1}{\sqrt{N}} \left( \begin{array}{c|c}
P^{(1)} & P^{(2)} \\
\hline
P^{(3)} & P^{(4)} \\
\hline
1_N & 0 \\
\hline
0 & 1_N
\end{array} \right),
\end{displaymath}
 where for $1 \leq i \leq 4$, the submatrices $(P^{(i)}_{kj})$ are $(N-1) \times N$-matrices given by $P^{(1)} := \big( \cos\frac{(2j-2) \pi k}{N} \big)_{kj}$, $P^{(2)} := \big( \cos\frac{(2j-1) \pi k}{N} \big)_{kj}$, $P^{(3)} := \big( \sin\frac{(2j-2) \pi k}{N} \big)_{kj}$, $P^{(4)} := \big( \sin\frac{(2j-1) \pi k}{N} \big)_{kj}$.

The inverse $(d_{b,a} \Phi)^{-1} = \Delta_1^{-1} P^T \Delta_2^{-1}$ at $(b,a) = (\b \cdot 1_N, \a \cdot 1_N)$ can then easily computed to be
\begin{equation} \label{dxybaba}
\left( \begin{array}{c|c|c|c}
&& -1 & 0 \\
\left( \sqrt{\frac{2\a}{N}} \, \l_k \cos \frac{(2j-2) \pi k}{N} \right)_{jk} & \left( \sqrt{\frac{2\a}{N}} \, \l_k \sin \frac{(2j-2) \pi k}{N} \right)_{jk} & \vdots & \vdots \\
&& -1 & 0 \\
\hline
&& 0 & 1 \\
\left( -\sqrt{\frac{\a}{2N}} \, \l_k \cos \frac{(2j-1) \pi k}{N} \right)_{jk} & \left( -\sqrt{\frac{\a}{2N}} \, \l_k \sin \frac{(2j-1) \pi k}{N} \right)_{jk} & \vdots & \vdots \\
&& 0 & 1
\end{array} \right),
\end{equation}
again with $1 \leq j \leq N$ and $1 \leq k \leq N-1$, but with $k$ and $j$ now being column and row indices, respectively.

The row vectors in the $2N \times 2N$-matrix (\ref{dbaxyc1c2}) are the gradients $\nba x_k$, $\nba y_k$ ($1 \leq k \leq N-1$) and $\nba C_i$ ($i=1,2$). In the sequel we will consider the restrictions of the coordinates $(x_k, y_k)_{1 \leq k \leq N-1}$ to the symplectic leaves $\Mba$. In order to obtain the gradients of these restricted functionals we need to take the orthogonal projections of the gradients $\nba x_k$, $\nba y_k$, $\nba C_i$ onto the tangent space $T_{b,a} \Mba$ of the leaf $\Mba$ at $(b,a)$. Note that $T_{b,a} \Mba (\cong \R^{2(N-1)})$ is the orthogonal complement of span$(\nba C_1, \nba C_2)$. In view of the formulas (\ref{c1grad})-(\ref{c2grad}) for the gradients of $C_1$ and $C_2$ one gets
\begin{equation} \label{tspcondgen}
T_{b,a} \Mba = \{ (\xi, \eta) \in \R^{2N} | \sum_{j=1}^N \xi_j = 0, \sum_{j=1}^N \frac{\eta_j}{a_j} = 0 \}.
\end{equation}
For $(b,a) = (\b 1_N, \a 1_N)$, the conditions in the set in (\ref{tspcondgen}) simply are $\sum_{j=1}^N \xi_j = 0$ and $\sum_{j=1}^N \eta_j = 0$. By formula (\ref{gradxnyn}), one sees that for $1 \leq k \leq N-1$,
\begin{displaymath}
  \sum_{j=1}^N \frac{\partial}{\partial b_j} (x_k + i y_k) = \frac{1}{\sqrt{2 \a N}} \frac{1}{\l_k} \sum_{j=1}^N e^{2\pi i(j-1)k/N} = 0,
\end{displaymath}
\begin{displaymath}
  \sum_{j=1}^N \frac{\partial}{\partial a_j} (x_k + i y_k) = -\frac{2}{\sqrt{2 \a N}} \frac{1}{\l_k} e^{i\pi k/N} \sum_{j=1}^N e^{2\pi i(j-1)k/N} = 0.
\end{displaymath}
Hence for $(b,a) = (\b 1_N, \a 1_N)$ the gradients $\nba x_k, \nba y_k$ are contained in $T_{b,a} \Mba$ for any $\1N1$, and there is no need to take projections.

\section{Linearized Birkhoff Coordinates}

For any given values of the Casimir functions, $C_1 = \b$, $C_2 = \a$, we wish to compute the first few coefficients of the Birkhoff normal form of the Toda Hamiltonian near the elliptic fixed point $(x,y) = (0,0)$. We recall that these coefficients are essentially unique, i.e. do not depend on the choice of the Birkhoff coordinates. A possible way of proceeding is to substitute $(b,a) = \Phi^{-1}(x,y,\b,\a)$ into the expression for the Toda Hamiltonian $H = \frac{1}{2} \sum_{j=1}^N b_j^2 + \sum_{j=1}^N a_j^2$ and then expand $\Phi^{-1}(x,y,\b,\a)$ at $(x,y) = (0,0)$. However it seems difficult to explicitly compute the terms of this expansion, except the first ones which we have computed in the last section - see formula (\ref{dxybaba}). We proceed differently and choose as the starting point of our computations the canonical coordinates $v_k$, $u_k$ introduced in section \ref{relcoord} rather than the \emph{non-canonical} variables $b_j$, $a_j$. When expressed in these coordinates, the Toda Hamiltonian takes the form $\Hba = \frac{N\b^2}{2} + H_u + \a^2 H_v$ where by (\ref{hbavu}),
\begin{eqnarray}
  H_u & = &\frac{1}{2} \left( u_1^2 + \sum_{l=1}^{N-2} (u_l - u_{l+1})^2 + u_{N-1}^2 \right), \label{hbeta}\\
  H_v & = & \sum_{l=1}^{N-1} e^{-v_l} + e^{\sum_{l=1}^{N-1} v_l}. \label{halpha}
\end{eqnarray}
Note that the Taylor expansion of $\Hba$ at $(v,u) = (0_{N-1}, 0_{N-1})$ is \emph{not} in Birkhoff normal form up to order $2$. In a first step we therefore want to choose a linear canonical transformation $(\xi_k, \eta_k)_{\1N1} \mapsto (v_k, u_k)_{\1N1}$ so that when expressed in the new variables $(\xi, \eta) = (\xi_k, \eta_k)_{\1N1}$, the Toda Hamiltonian is in Birkhoff normal form up to order $2$. Consider the composition
\begin{equation} \label{obadef}
\begin{array}{cccccc}
  \Oba: & \R^{2N-2} (\cong \Pba) & \to & \Mba & \to & \R^{2N-2} \\
& (x_k, y_k)_{\1N1} & \mapsto & (b_j, a_j)_{1 \leq j \leq N} & \mapsto & (v_k, u_k)_{\1N1} \end{array}
\end{equation}
of the inverse of the Birkhoff map $(\Phi|_{\Mba})^{-1}: \R^{2N-2} \to \Mba$ with the coordinate transformation defined in (\ref{vuba}). Then $\Oba$ is a canonical real analytic transformation as both $(x,y)$ and $(v,u)$ are canonical coordiantes for $\Mba$. Its Jacobian
\begin{equation} \label{rbadef}
  \Rba: \R^{2N-2} \to \R^{2N-2}, \quad (\xi, \eta) \mapsto (v,u) = d_{x,y} \Oba|_{(x,y)=(0,0)} (\xi, \eta)
\end{equation}
at $(x,y) = (0,0)$ is a linear transformation with the desired properties. We will compute $\Rba$ as a composition of the Jacobian of $(x,y,\b,\a) \mapsto (b,a)$ at $(x,y) = (0,0)$ (with $\b$, $\a$ fixed) and the one of $(b,a) \mapsto (v,u)$ at $(b,a) = (\b 1_N, \a 1_N)$.

It is convenient to use complex notation for $\xi_k$, $\eta_k$ ($\1N1$),
\begin{displaymath}
  \z_k := \frac{1}{\sqrt{2}} (\xi_k - i \eta_k), \quad \z_{-k} := \frac{1}{\sqrt{2}} (\xi_k + i \eta_k),
\end{displaymath}
where the sign in the definition of $\z_k$ is chosen so that $d\z_k \wedge d\z_{-k} = i d\xi_k \wedge d\eta_k$. The vector $\z = (\z_k)_{1 \leq |k| \leq N-1}$ is an element in the space
\begin{equation} \label{zspace}
  \mathcal{Z} := \{ \z = (\z_k)_{1 \leq |k| \leq N-1} \in \C^{2N-2} : \z_{-k} = \overline{\z_k} \quad \forall \; \1N1 \}.
\end{equation}
The components of $\z$ satisfy the identity
\begin{equation} \label{realcomplex}
  e^{i \pi j k/N} \z_k + e^{-i \pi j k/N} \z_{-k} = \sqrt{2} \, \left( \cos \left( \frac{j \pi k}{N} \right) \xi_k + \sin \left( \frac{j \pi k}{N} \right) \eta_k \right).
\end{equation}

As for any $\1N1$, $u_k$ is a linear function of the $b_j$'s by (\ref{vuba}) one has 
\begin{displaymath}
  u_k = -\frac{k}{N} \sum_{j=1}^N b_j(\z) + \sum_{j=1}^k b_j(\z) \quad (\1N1)
\end{displaymath}
where $b_j(\z)$ ($1 \leq j \leq N$) can be computed using formula (\ref{realcomplex}) and the Jacobian of $(\Phi|_{\Mba})^{-1}$ at $(x,y) = (0,0)$ obtained in (\ref{dxybaba}) to get, for $1 \leq j \leq N$,
\begin{eqnarray*}
  b_j(\z) & = & \sqrt{\a / N} \sum_{k=1}^{N-1} \l_k \left( e^{2\pi i(j-1)k/N} \z_k + e^{-2\pi i(j-1)k/N} \z_{-k} \right) \\
& = & \sqrt{\a / N} \sum_{1 \leq |k| \leq N-1} \l_k e^{2\pi i(j-1)k/N} \z_k,
\end{eqnarray*}
where we used that $\l_k = \l_{-k}$ for any $\1N1$. Note that $\sum_{j=1}^N b_j(\z) = 0$ as $\sum_{j=1}^N e^{2\pi i (j-1)k/N} = 0$. To compute $H_u = \frac{1}{2} (u_1^2 + \sum_{l=1}^{N-2} (u_l - u_{l+1})^2 + u_{N-1}^2)$ we use
\begin{eqnarray}
  u_1(\z) = b_1(\z) & = & \sqrt{\a / N} \sum_{1 \leq |k| \leq N-1} \l_k \z_k, \label{transfu1} \\
u_l(\z)  - u_{l+1}(\z) = -b_{l+1}(\z)  & = & -\sqrt{\a / N} \sum_{1 \leq |k| \leq N-1} \l_k e^{2\pi i lk/N} \z_k, \label{transfu2} \\
-u_{N-1}(\z) = -\sum_{j=1}^{N-1} b_j(\z) = b_N(\z) & = & \sqrt{\a / N} \sum_{1 \leq |k| \leq N-1} \!\!\!\! \l_k e^{2 \pi i (N-1) k/N} \z_k. \label{transfu3}
\end{eqnarray}
with $1 \leq l \leq  N-2$ in (\ref{transfu2}). Thus
\begin{eqnarray*}
  H_u(\z) & = & \frac{1}{2} \frac{\a}{N} \sum_{l=0}^{N-1} \left( \sum_{1 \leq |k| \leq N-1} \l_k e^{2\pi i lk/N} \z_k \right)^2 \\
& = & \frac{1}{2} \frac{\a}{N} \sum_{1 \leq |k|, |k'| \leq N-1} \l_k \l_{k'} \left( \sum_{l=0}^{N-1} e^{2\pi i l(k+k')/N} \right) \z_k \z_{k'}.
\end{eqnarray*}
Using that $\sum_{l=0}^{N-1} e^{2\pi i lk/N} = N \delta_{k0}$ for any $0 \leq |k| \leq N-1$, one gets an expression which is quadratic in the $\z$-variables,
\begin{equation}
  H_u(\z) = \a \sum_{k=1}^{N-1} \l_k^2 \z_k \z_{-k}.
\end{equation}

Now let us turn to $H_v$ given by formula (\ref{halpha}). As $(b,a) = (\b 1_N, \a 1_N)$ we have by (\ref{dvkaj}),
\begin{displaymath}
 v_k = -\frac{2}{\a} a_k(\z) + \frac{2}{N\a} \sum_{j=1}^N a_j(\z)
\end{displaymath}
where $a_j(\z)$ ($1 \leq j \leq N$) can be computed using formula (\ref{realcomplex}) and the Jacobian $d_{x,y,\b,\a} \Phi^{-1}$ at $(x,y) = (0,0)$ obtained in (\ref{dxybaba}) to get
\begin{eqnarray*}
  a_j(\z) & = & -\frac{1}{2} \sqrt{\a / N} \sum_{k=1}^{N-1} \l_k \left( e^{i\pi (2j-1)k/N} \z_k + e^{-i\pi (2j-1)k/N} \z_{-k} \right) \\
& = & -\frac{1}{2} \sqrt{\a / N} \sum_{1 \leq |k| \leq N-1} \l_k e^{i\pi (2j-1)k/N} \z_k.
\end{eqnarray*}
Again we have that $\sum_{j=1}^N a_j(\z) = 0$ as for any $1 \leq |k| \leq N-1$,
\begin{displaymath}
  \sum_{j=1}^N e^{i\pi (2j-1)k/N} = e^{-i\pi k/N} \left( \sum_{j=1}^N e^{2\pi i jk/N} \right) = 0.
\end{displaymath}
Hence, for $1 \leq l \leq N-1$,
\begin{equation} \label{vlzk}
  v_l = \frac{1}{\sqrt{\a N}} \sum_{1 \leq |k| \leq N-1} \l_k e^{2\pi i lk/N} e^{-i\pi k/N} \z_k.
\end{equation}
Define $v_0$ by the expression on the right hand side of (\ref{vlzk}) evaluated at $l=0$. Note that
\begin{equation} \label{sumvl}
  \sum_{l=0}^{N-1} v_l = \frac{1}{\sqrt{\a N}} \sum_{1 \leq |k| \leq N-1} \l_k \z_k e^{-i\pi k/N} \left( \sum_{l=1}^{N-1} e^{2\pi i lk/N} \right) = 0.
\end{equation}
Hence $\sum_{l=1}^{N-1} v_l = -v_0$ and therefore
\begin{displaymath}
  H_v = \sum_{l=0}^{N-1} e^{-v_l}.
\end{displaymath}
Now we expand $H_v$ at $v=0$ up to order $4$,
\begin{equation} \label{havkexp}
  H_v = N - \sum_{l=0}^{N-1} v_l + \frac{1}{2} \sum_{l=0}^{N-1} v_l^2 - \frac{1}{3!} \sum_{l=0}^{N-1} v_l^3 + \frac{1}{4!} \sum_{l=0}^{N-1} v_l^4 + O(|v|^5).
\end{equation}
By (\ref{sumvl}), $\sum_{l=0}^{N-1} v_l = 0$. Substituting the expressions (\ref{vlzk}) for $v_l$ in the quadratic term in the expansion (\ref{havkexp}) we get
\begin{eqnarray*}
  \sum_{l=0}^{N-1} v_l^2 & = & \frac{1}{\a N} \sum_{1 \leq |k|, |k'| \leq N-1} \!\! \l_k \l_{k'} \left( \sum_{l=0}^{N-1} e^{2\pi i l(k+k')/N} \right) e^{-i\pi (k+k')/N} \z_k \z_{k'} \\
& = & \frac{2}{\a} \sum_{k=1}^{N-1} \l_k^2 \z_k \z_{-k},
\end{eqnarray*}
where we again used that $\l_k = \l_{-k}$ and that $\sum_{l=0}^{N-1} e^{2\pi i lk/N} = N \delta_{k0}$ for any $0 \leq |k| \leq N-1$. The terms of third and fourth order in $H_v$ are treated similarly. Combining the above computations leads to
\begin{prop}
Let $\b \in \R$ and $\a>0$. Then
\begin{equation} \label{hbarbazeta}
  \hrba(\z) = G_0 + \a G_2 + \sqrt\a G_3 + G_4 + O(\z^5)
\end{equation}
where $\Rba$ is the linear canonical transformation introduced in (\ref{rbadef}) and $G_i$ ($0 \leq i \leq 4$) are given by
\begin{eqnarray}
G_0 & := & \frac{N\b^2}{2} + N\a^2, \label{g0def} \\
G_2 & := & 2 \sum_{k=1}^{N-1} \l_k^2 \z_k \z_{-k}, \label{g2lambdadef} \\
G_3 & := & -\frac{1}{6 \sqrt N}  \sum_{1 \leq |k|, |k'|, |k''| \leq N-1 \atop k+k'+k'' \equiv 0 \, mod \, N} (-1)^{(k+k'+k'')/N} \l_k \l_{k'} \l_{k''} \z_k \z_{k'} \z_{k''}, \label{g3lambdadef} \\
G_4 & := & \frac{1}{24N} \!\! \sum_{1 \leq |k|, |k'|, |k''|, |k'''| \leq N-1 \atop k+k'+k''+k''' \equiv 0 \, mod  \, N} \!\!\!\!\!\!\!\!\!\!\!\!\!\!\!\! (-1)^{(k\!+\!k'\!+\!k''\!+\!k''')/N} \l_k \l_{k'} \l_{k''} \l_{k'''} \z_k \z_{k'} \z_{k''} \z_{k'''}. \label{g4lambdadef}
\end{eqnarray}
\end{prop}

Note that $\hrba(\z)$ depends on $\b$ only through the constant term $\frac{N\b^2}{2}$ and that it is in Birkhoff normal form up to order $2$.

To finish this section let us express the Birkhoff coordinates $(x,y)$ in terms of the coordinates $(\xi, \eta)$ near the origin. The two coordinate systems are related by
\begin{displaymath}
  (x,y) = (\Oba)^{-1} d_{0,0} \Oba (\xi, \eta)
\end{displaymath}
where we used that $\Rba = d_{0,0} \Oba$. Hence
\begin{equation} \label{xkykxiketak}
  (x_k, y_k) = (\xi_k, \eta_k) + O(|(\xi, \eta)|^2) \quad \forall \, \1N1
\end{equation}
and
\begin{displaymath}
  I_k = \frac{x_k^2 + y_k^2}{2} = \frac{\xi_k^2 + \eta_k^2}{2} + O(|(\xi, \eta)|^3) \quad \forall \, \1N1.
\end{displaymath}

Denote by $\cbao$ the Poisson algebra of $\hrba$, i.e. the space of germs of real
analytic functions $F(\xi, \eta) = \sum_{|\g| + |\d| \geq 2} f_{\g \d} \xi^\g
\eta^\d$ such that $\{ F, I_k \} = 0$ for any $\1N1$. In view of (\ref{xkykxiketak}) we say that $\cbao$ is \emph{non-resonant}. The following result is then well known (see e.g. \cite{kapo}, Appendix G).

\begin{cor} \label{bnftransformcor}
For any $\b \in \R$, $\a>0$, and $m \geq 3$, there exists a (germ of a) real analytic canonical transformation of the form \idplgeq given by $X^t_{F_m}|_{t=1}$, where $X^t_{F_m}$ is the flow of the Hamiltonian vector field associated to the Hamiltonian
\begin{displaymath}
  F_m(\xi, \eta) = \sum_{3 \leq |\g| + |\d| \leq m} f_{\g \d} \xi^\g \eta^\d,
\end{displaymath}
such that $F \circ X^t_{F_m}|_{t=1}$ is in Birkhoff normal form up to order $m$ for \emph{any} $F$ in $\cbao$.
\end{cor}

\section{Proof of Theorem \ref{bnftodatheorem}}

We now transform the Hamiltonian $\hrba(\z)$ into its Birkhoff normal form up to order $4$ by a standard procedure - see e.g. section $14$ in \cite{kapo}. The phase space $\mathcal{Z}$, defined in (\ref{zspace}), is endowed with the Poisson bracket
\begin{displaymath}
  \{ F,G \} = i \sum_{1 \leq |k| \leq N-1} \sigma_k \frac{\partial F}{\partial \z_k} \frac{\partial G}{\partial \z_{-k}},
\end{displaymath}
where $\sigma_k = \textrm{sgn} \, (k)$ is the sign of $k$. The Hamiltonian vector field $X_F$ associated to the Hamiltonian $F$ is then given by $X_F = i \sum_{1 \leq |k| \leq N-1} \sigma_k \frac{\partial H}{\partial \z_{-k}} \frac{\partial}{\partial \z_k}$. With a first canonical transformation we want to eliminate the third order term $G_3$ in the expansion (\ref{hbarbazeta}) of $\hrba(\z)$. We construct such a canonical transformation on the phase space $\mathcal{Z}$ as the time-$1$-map $\Psi_1 := X^t_F|_{t=1}$ of the flow $X^t_F$ of a real analytic Hamiltonian $F := \a^{-1/2} F_3$ which is a homogeneous polynomial in $\z_k$ ($1 \leq |k| \leq N-1$) of degree $3$ and solves the following homological equation
\begin{equation} \label{homeqn}
  \{ \a G_2, \a^{-\frac{1}{2}} F_3 \} + \a^\frac{1}{2} G_3 = 0.
\end{equation}
To simplify notation we momentarily write $H$ instead of $\hrba$. Assuming for the moment that (\ref{homeqn}) can be solved and that $X^t_F$ is defined for $0 \leq t \leq 1$ in some neighbourhood of the origin in $\mathcal{Z}$ we can use Taylor's formula to expand $H \circ X^t_F$ around $t=0$,
\begin{eqnarray}
  H \circ X^t_F & = & H \circ X^0_F + \int_0^t \frac{d}{ds}(H \circ X^s_F) ds = H + \int_0^t \{ H, F \} \circ X^s_F \, ds \nonumber\\
& = & H + t \, \{ H, F \} + \int_0^t (t-s) \{ \{ H, F \}, F \} \circ X^s_F \, ds. \label{hbataylort}
\end{eqnarray}
When evaluating this expression at $t=1$, one gets
\begin{eqnarray*}
  H \circ \Psi_1 & = & G_0 + \a G_2 + \{ \a G_2, F \} + \int_0^1 (1-t) \{ \{ \a G_2, F \}, F \} \circ X^t_F dt \\
&& \quad + \sqrt\a G_3 + \int_0^1 \{ \sqrt\a G_3, F \} \circ X^t_F \, dt + G_4 + O(\z^5).
\end{eqnarray*}
Using that $\{ \a G_2, F \} + \sqrt\a G_3 = 0$, the latter expression simplifies and we get
\begin{displaymath}
  H \circ \Psi_1 = G_0 + \a G_2 + \int_0^1 t \, \{ \sqrt\a G_3, F \} \circ X^t_F \, dt + G_4 + O(\z^5).
\end{displaymath}
Integrating by parts once more and taking into account that $F = \a^{-1/2} F_3$ is homogeneous of degree $3$ one obtains, in view of (\ref{hbataylort}),
\begin{equation} \label{H3}
  \hrba \circ \Psi_1 = G_0 + \a G_2 + G_4 + \frac{1}{2} \{ G_3, F_3 \} + O(\z^5).
\end{equation}
Note that $\{ G_3, F_3 \}$ is homogeneous of order $4$. Hence our first step is achieved. It remains to solve (\ref{homeqn}).

By Corollary \ref{bnftransformcor} with $m=3$ there exists a polynomial homogeneous of degree $3$,
\begin{displaymath}
  W_3 = \sum_{1 \leq |k|, |k'|, |k''| \leq N-1} W_{kk'k''}^{(3)} \z_k \z_{k'} \z_{k''},
\end{displaymath}
(with $W_{kk'k''}^{(3)}$ invariant under permutations of $k$, $k'$, $k''$) so that the time-$1$-map $X_W^t|_{t=1}$ of the flow $X_W^t$ corresponding to the Hamiltonian $W = \a^{-1/2} W_3$ brings any Hamiltonian in $\cbao$ into Birkhoff normal form up to order $3$. In particular, the identity $\{ G_2, W_3 \} + G_3 = 0$ is satisfied. Note that
\begin{eqnarray*}
\{ G_2, W_3 \} & = & i \sum_{1 \leq |k| \leq N-1} 2 \sigma_k \l_k^2 \z_{-k} \frac{\partial W_3}{\partial \z_{-k}}\\
& = & -i \sum_{1 \leq |k|, |k'|, |k''| \leq N-1} (s_k + s_{k'} + s_{k''}) W^{(3)}_{kk'k''} \z_k \z_{k'} \z_{k''}
\end{eqnarray*}
as $s_k = 2 \sin \frac{k \pi}{N} = 2 \sigma_k \l_k^2$, and it follows that
\begin{equation} \label{swkkkgkkk}
  (s_k + s_{k'} + s_{k''}) i W_{kk'k''}^{(3)} = G_{kk'k''}^{(3)}
\end{equation}
where $G_{kk'k''}^{(3)}$ are the coefficients of $G_3$ defined by (\ref{g3lambdadef}),
\begin{displaymath}
  G^{(3)}_{kk'k''} = \left\{  \begin{array}{ll}
-1/(6 \sqrt N) (-1)^{(k+k'+k'')/N} \l_k \l_{k'} \l_{k''} & \qquad \textrm{for} \; k\!+\!k'\!+\!k'' \equiv 0 \, \textrm{mod} \, N \\
0 & \qquad \textrm{otherwise.}
\end{array} \right.
\end{displaymath}
As $G^{(3)}_{kk'k''} \neq 0$ iff $k+k'+k'' \equiv 0$ mod $N$ it follows\footnote{See \cite{rink01} (cf. also \cite{ahtk5}) for a direct proof of this nonresonance condition.} from (\ref{swkkkgkkk}) that for any triple $(k,k',k'')$ with $k+k'+k'' \equiv 0$ mod $N$,
\begin{displaymath}
  s_k + s_{k'} + s_{k''} \neq 0
\end{displaymath}
and
\begin{displaymath}
i W^{(3)}_{kk'k''} = \frac{G^{(3)}_{kk'k''}}{s_k + s_{k'} + s_{k''}}.
\end{displaymath}
Now define $F_3 = \sum_{1 \leq |k|, |k'|, |k''| \leq N-1} F^{(3)}_{kk'k''} \z_k \z_{k'} \z_{k''}$ and $F = \a^{-1/2} F_3$ by
\begin{equation} \label{fkkkformula}
  F^{(3)}_{kk'k''} = \left\{  \begin{array}{ll}
W^{(3)}_{kk'k''} & \qquad \textrm{for} \; k+k'+k'' \equiv 0 \, \textrm{mod} \, N \\
0 & \qquad \textrm{otherwise.}
\end{array} \right.
\end{equation}
Then clearly $\{ G_2, F_3 \} + G_3 = 0$ and hence by the considerations above $\Psi_1 := X_F^t|_{t=1}$ has the property that $\hrba \circ \Psi_1$ satisfies identity (\ref{H3}).

Now let us investigate the $4$th order term $G_4 + \frac{1}{2} \{ G_3, F_3 \}$ in (\ref{H3}). We decompose this sum into its contribution to the Birkhoff normal form and the rest, to be transformed away in a moment. Let us first compute $\{ G_3, F_3 \}$ in a more explicit form. By (\ref{g3lambdadef}) and (\ref{fkkkformula}),
\begin{eqnarray*}
  \{ G_3, F_3 \} & = & i \sum_{1 \leq |k| \leq N-1} \sigma_k \frac{\partial G_3}{\partial \z_k} \frac{\partial F_3}{\partial \z_{-k}}\\
& = & \frac{1}{36N} \sum_{1 \leq |k| \leq N-1} \sigma_k \left( 3 \!\!\!\!\!\! \sum_{1 \leq |l|, |m| \leq N-1, \atop l+m = -k+rN} \!\!\!\! \, (-1)^r \l_k \l_l \l_m \z_l \z_m \right) \\
&& \qquad \qquad \cdot \left( 3 \!\!\!\!\!\! \sum_{1 \leq |l'|, |m'| \leq N-1, \atop l'+m' = k+r'N} \!\!\!\! (-1)^{r'} \frac{\l_k \l_{l'} \l_{m'}}{s_{-k} \!+\! s_{l'} \!+\! s_{m'}} \z_{l'} \z_{m'} \right) \\
& = & \frac{1}{8N} \!\!\!\! \sum_{1 \leq |k| \leq N-1} \sum_{1 \leq |l|, |m|, |l'|, |m'| \leq N-1 \atop {l+m-rN = -k \atop l'+m'-r'N = k}} \!\!\!\!\!\! (-1)^{r+r'} \frac{s_k \l_l \l_m \l_{l'} \l_{m'}}{s_{-k} + s_{l'} + s_{m'}} \z_l \z_m \z_{l'} \z_{m'},
\end{eqnarray*}
where for the latter equality we used again that $2 \sigma_k \l_k^2 = s_k$. Setting
\begin{displaymath}
\e_{lml'm'} = \frac{l+m+l'+m'}{N}
\end{displaymath}
and using that $s_{-k} = -s_k$ one then gets
\begin{eqnarray*}
  \{ G_3, F_3 \} & = & \frac{1}{8N} \!\!\!\! \sum_{1 \leq |k| \leq N-1} \sum_{l+m \equiv -k \, mod \, N \atop {l'+m'\equiv k \, mod \, N \atop \ldots}} \!\!\!\! (-1)^{\e_{lml'm'}} \frac{\l_l \l_m \l_{l'} \l_{m'}}{-1 + (s_{l'} + s_{m'})/s_k} \z_l \z_m \z_{l'} \z_{m'} \\
& = & \frac{1}{8N} \sum_{k=1}^{N-1} \sum_{l+m \equiv -k \, mod \, N \atop {l'+m' \equiv k \, mod \, N \atop \ldots}} \!\!\!\! (-1)^{\e_{lml'm'}} \frac{\l_l \l_m \l_{l'} \l_{m'}}{-1 + (s_{l'} + s_{m'})/s_k} \z_l \z_m \z_{l'} \z_{m'} \\
&& + \frac{1}{8N} \sum_{k=1}^{N-1} \sum_{l+m \equiv k \, mod \, N \atop {l'+m'\equiv -k \, mod \, N \atop \ldots}} \!\!\!\! (-1)^{\e_{lml'm'}} \frac{\l_l \l_m \l_{l'} \l_{m'}}{-1 - (s_{l'} + s_{m'})/s_k} \z_l \z_m \z_{l'} \z_{m'} \\
& = & \frac{1}{8N} \sum_{k=1}^{N-1} \sum_{l+m \equiv -k \, mod \, N \atop {l'+m' \equiv k \, mod \, N \atop \ldots}} \!\!\! \left( \frac{1}{-1 + (s_{l'} \!+\! s_{m'})/s_k} + \frac{1}{-1 - (s_l \!+\! s_m)/s_k} \right) \\ 
&& \qquad \qquad \qquad \qquad \cdot \, (-1)^{\e_{lml'm'}} \l_l \l_m \l_{l'} \l_{m'} \z_l \z_m \z_{l'} \z_{m'}
\end{eqnarray*}
where $\ldots$ stand for $1 \leq |l|, |m|, |l'|, |m'| \leq N-1$. Note that for $k = l'+m'+r'N$ with $\1N1$ and $r' \in \Z$ we have
\begin{displaymath}
  s_k = |s_{l'+m'}|.
\end{displaymath}
Hence
\begin{equation} \label{g3f3}
  \frac{1}{2} \{ G_3, F_3 \} = \frac{1}{24N} \!\!\!\!\!\!\!\!\!\!\!\! \sum_{ \atop {l+m+l'+m' \equiv 0 \, mod \, N \atop \ldots}} \!\!\!\!\!\!\!\!\!\!\!\! c_{lml'm'} (-1)^{\e_{lml'm'}} \l_l \l_m \l_{l'} \l_{m'} \z_l \z_m \z_{l'} \z_{m'}
\end{equation}
where\footnote{To keep the formulas as simple as possible we have not symmetrized the coefficients $c_{lml'm'}.$}
\begin{equation} \label{cklmndef}
  c_{lml'm'} = \left\{ \begin{array}{ll}
 \frac{3}{2} \left( \frac{1}{-1 + \frac{s_{l'} + s_{m'}}{|s_{l'+m'}|}} - \frac{1}{1 + \frac{s_l + s_m}{|s_{l+m}|}} \right) & \left\{ \begin{array}{l} 1 \leq |l|, |m|, |l'|, |m'| \leq N-1 \\ l+m \not \equiv 0 \; \textrm{mod} \, N \\ l+m+l'+m' \equiv 0 \; \textrm{mod} \, N \end{array} \right. \\
&\\
0 & \quad \textrm{otherwise} \end{array} \right.
\end{equation}

Combined with formula (\ref{g4lambdadef}) for $G_4$, the sum $G_4 + \frac{1}{2} \{ G_3, F_3 \}$ equals
\begin{equation} \label{4thordercomplete}
\frac{1}{24N} \!\!\!\!\!\!\!\!\!\!\!\!\!\! \sum_{{} \atop { {} \atop {1 \leq |k|, |k'|, |k''|, |k'''| \leq N-1 \atop {k+k'+k''+k''' \equiv 0 \, mod \, N}}}} \!\!\!\!\!\!\!\!\!\!\!\!\!\! (-1)^{\e_{kk'k''k'''}} (1 + c_{kk'k''k'''}) \cdot \l_k \l_{k'} \l_{k''} \l_{k'''} \z_k \z_{k'} \z_{k''} \z_{k'''}.
\end{equation}
We now decompose (\ref{4thordercomplete}) into its contribution $\pi_N \!\! \left( G_4 \!+\! \frac{1}{2} \{ G_3, F_3 \} \right)$ to the Birkhoff normal form of $\hrba$ and the rest.

\begin{lemma}
The normal form part of $G_4 + \frac{1}{2} \{ G_3, F_3 \}$ is given by
\begin{equation} \label{ping4ping3f3}
\pi_N \left( G_4 + \frac{1}{2} \{ G_3, F_3 \} \right) = \frac{1}{4N} \sum_{l=1}^{N-1} |\z_l|^4.
\end{equation}
\end{lemma}

\begin{proof}
The indices $k,k',k'',k'''$ of the terms in (\ref{4thordercomplete}) contributing to the normal form satisfy the condition
\begin{equation} \label{nfcond}
\{ k,k',k'',k''' \} = \{ l,-l,m,-m \}
\end{equation}
with $1 \leq l \leq m \leq N-1$. In the case $l=m$, $\{ l,-l,l,-l \}$ is viewed as a set-like object whose two elements $l$ and $-l$ each have multiplicity two.

We investigate $\pi_N(G_4)$ and $\pi_N(\frac{1}{2} \{ G_3, F_3 \})$ separately. Let us start with $G_4$. We distinguish the cases $l=m$ and $l \neq m$ in (\ref{nfcond}). For $l=m$, there are ${4 \choose 2} = 6$ distinct combinations of indices $(k,k',k'',k''')$ satisfying (\ref{nfcond}), whereas for $l \neq m$, there are $4! = 24$ (automatically distinct) permutations of $(l,m,-l,-m)$. Hence we have
\begin{eqnarray}
  \pi_N(G_4) & = & \frac{1}{24N} \left( 6 \sum_{l=1}^{N-1} \l_l^4 |\z_l|^4 + 24 \!\!\!\! \sum_{1 \leq l < m \leq N-1} \!\!\!\! \l_l^2 \l_m^2 |\z_l|^2 |\z_m|^2 \right) \nonumber\\
& = & \frac{1}{4N} \left( \sum_{l=1}^{N-1} \l_l^4 |\z_l|^4 + 4 \!\!\!\! \sum_{1 \leq l < m \leq N-1} \!\!\!\! \l_l^2 \l_m^2 |\z_l|^2 |\z_m|^2 \right). \label{ping4formula}
\end{eqnarray}

Now let us compute $\pi_N(\frac{1}{2} \{ G_3, F_3 \})$. We have to single out the matches of (\ref{nfcond}) for which in addition $c_{kk'k''k'''} \neq 0$, i.e.
\begin{displaymath}
k+k' \neq 0 \, \textrm{mod} \, N, \; k+k'+k''+k''' \equiv 0 \, \textrm{mod} \, N
\end{displaymath}
To fulfill (\ref{nfcond}), there are therefore the two possibilities
\begin{equation} \label{kmlnknlm}
\left. \begin{array}{rl}
k+k'' & = 0 \\
k'+k''' & = 0 \end{array} \right. \qquad \qquad \textrm{or} \qquad \qquad \left. \begin{array}{rl}
k+k''' & = 0 \\
k'+k'' & = 0 \end{array} \right..
\end{equation}
In both cases, we have $s_{k''} + s_{k'''} = -(s_k + s_{k'})$, and therefore (\ref{cklmndef}) reduces to
\begin{equation} \label{cklmnnfterms}
  c_{kk'k''k'''} = \frac{-3 |s_{k+k'}|}{|s_{k+k'}| + s_k + s_{k'}}.
\end{equation}
Note that (\ref{cklmnnfterms}) remains valid for $k+k' = N$, since in this case $s_{k+k'} = 0$ and $s_k + s_{k'} > 0$ as $k$ and $k'$ must satisfy $1 \leq k,k' \leq N-1$, but not for $k+k' = 0$, since in this case $|s_{k+k'}| + s_k + s_{k'} = 0$.

We first compute the diagonal part of $\pi_N \big( \frac{1}{2} \{ G_3, F_3 \} \big)$. In this case, the two possibilities in (\ref{kmlnknlm}) coincide and the solutions are
\begin{equation} \label{diaggf3coeff}
(k,k',k'',k''') = \left\{ \begin{array}{cccc} 
  (l, & l, & -l, & -l) \\
  (-l, & -l, & l, & l) \end{array} \right..
\end{equation}
The sum of the coefficients $c_{kk'k''k'''}$ for the two cases listed in (\ref{diaggf3coeff}) is
\begin{displaymath}
  c_{l,l,-l,-l} + c_{-l,-l,l,l} = -3 |s_{2l}| \left( \frac{1}{|s_{2l}| \!+\! 2 s_l} + \frac{1}{|s_{2l}| \!-\! 2 s_l} \right) = \frac{-6 s_{2l}^2}{s_{2l}^2 \!-\! 4 s_l^2} = 6 \cot^2 \frac{l \pi}{N}.
\end{displaymath}

We now turn to the off-diagonal part of $\pi_N \big( \frac{1}{2} \{ G_3, F_3 \} \big)$. The matches of (\ref{nfcond}), (\ref{kmlnknlm}) for given $\{ l, m \} \subseteq \{ 1, \ldots, N-1 \}$ with $l < m$, $(k,k') = (\pm l, \pm m)$ and $(k'',k''') = (\pm l, \pm m)$ are
\begin{equation} \label{offdiaggf3coeff}
(k,k',k'',k''') = \left\{ \begin{array}{cccc} 
  (l, & m, & -l, & -m) \\
  (l, & -m, & -l, & m) \\
  (-l, & m, & l, & -m) \\
  (-l, & -m, & l, & m) \end{array} \right..
\end{equation}
The remaining matches are obtained from (\ref{offdiaggf3coeff}) by permuting the first and second or the third and fourth columns on the right hand side of (\ref{offdiaggf3coeff}), bringing the total of all matches to $16 = 4 \cdot 4$. Note that by formula (\ref{cklmnnfterms}), these permutations leave the value of the coefficients $c_{kk'k''k'''}$ invariant. Taking the sum of the coefficients $c_{kk'k''k'''}$ for all the matches, we obtain from (\ref{cklmnnfterms})
\begin{eqnarray*}
 4 (c_{l,m,-l,-m} & + & c_{l,-m,-l,m} + c_{-l,m,l,-m} + c_{-l,-m,l,m}) \\
& = & -12 \Bigg( \frac{|s_{l+m}|}{|s_{l+m}| + s_l + s_m} + \frac{|s_{l-m}|}{|s_{l-m}| + s_l - s_m} + \\
&& \quad \frac{|s_{l-m}|}{|s_{l-m}| - s_l + s_m} + \frac{|s_{l+m}|}{|s_{l+m}| - s_l - s_m} \Bigg) \\
& = & -24 \left( \frac{s_{l-m}^2}{s_{l-m}^2 - (s_l - s_m)^2} + \frac{s_{l+m}^2}{s_{l+m}^2 - (s_l + s_m)^2} \right) \\
& = & \frac{-24 (2 s_{l-m}^2 s_{l+m}^2 - s_{l-m}^2 (s_l + s_m)^2 - s_{l+m}^2 (s_l - s_m)^2)}{s_{l \!-\! m}^2 s_{l \!+\! m}^2 + (s_l \!-\! s_m)^2 (s_l \!+\! s_m)^2 \!-\! s_{l \!-\! m}^2 (s_l \!+\! s_m)^2 \!-\! s_{l \!+\! m}^2 (s_l \!-\! s_m)^2} \\
& = & -24,
\end{eqnarray*}
since $s_{l-m}^2 s_{l+m}^2 = (s_l-s_m)^2 (s_l+s_m)^2$. Collecting terms, we thus have
\begin{eqnarray}
  \pi_N \left( \frac{1}{2} \{ G_3, F_3 \} \right) & = & \frac{1}{24N} \left( \sum_{l=1}^{N-1} 6 \cos^2 \frac{\pi l}{N} |\z_l|^4 - 24 \!\!\!\!\!\! \sum_{1 \leq l < m \leq N-1} \!\!\!\!\!\! \l_l^2 \l_m^2 |\z_l|^2 |\z_m|^2 \right) \nonumber\\
& = & \frac{1}{4N} \left( \sum_{l=1}^{N-1} (1 - \l_l^4) |\z_l|^4 - 4 \!\!\!\!\!\! \sum_{1 \leq l < m \leq N-1} \!\!\!\!\!\! \l_l^2 \l_m^2 |\z_l|^2 |\z_m|^2 \right). \label{ping3f3formula}
\end{eqnarray}
Adding up (\ref{ping4formula}) and (\ref{ping3f3formula}), we obtain (\ref{ping4ping3f3}).
\end{proof}

In a next step we want to remove $(1 - \pi_N) \left( G_4 + \frac{1}{2} \{ G_3, F_3 \} \right)$ from the Hamiltonian by a second coordinate transformation $\Psi_2$. In view of Corollary \ref{bnftransformcor} with $m=4$, there exists a canonical transformation $\Psi_2$ of the form \idplgeq \, so that for any Hamiltonian $F$ in the Poisson algebra $\Psi_1^* \cbao$, $F \circ \Psi_2$ is in Birkhoff normal form up to order $4$. We have proved the following

\begin{prop} \label{localtheorem}
Let $\b \in \R$ and $\a>0$ be given. The real analytic symplectic coordinate transformation $\z = \Xi(z) = \Psi_1 \circ \Psi_2 (z)$ defined in a neighborhood of the origin in $\mathcal{Z}$, transforms the Hamiltonian $\hrba$ into its Birkhoff normal form up to order $4$. More precisely,
\begin{equation} \label{hamtaylorwk}
  \hrba \circ \Xi = G_0 + \a G_2 + \pi_N \left( G_4 + \frac{1}{2} \{ G_3, F_3 \} \right) + O(z^5),
\end{equation}
with $G_0$, $G_2$ and $\pi_N(G_4 + \frac{1}{2} \{ G_3, F_3 \})$ defined by (\ref{g0def}), (\ref{g2lambdadef}), and (\ref{ping4ping3f3}).
\end{prop}

\begin{proof}[Proof of Theorem \ref{bnftodatheorem}]
The map $\Oba: \Pba \to \R^{2N-2}$ introduced in (\ref{obadef}) is a canonical real analytic diffeomorphism so that $\Hba \circ \Oba = (\hrba) \circ (\Rba^{-1} \circ \Oba)$ is in Birkhoff normal form. By the definition (\ref{rbadef}), the map $\Rba^{-1} \circ \Oba$ is canonical and, near the origin, of the form \idplgeq. The canonical transformation $\Xi$ of Proposition \ref{localtheorem} is also of the form \idplgeq and brings $\hrba$ \emph{locally} around the origin into Birkhoff normal form up to order $4$. Hence the transformations $\Rba^{-1} \circ \Oba$ and $\Xi$ differ, near the origin, by a transformation of the form \idplgeq. By the uniqueness of the Birkhoff normal form, the expansions of the Toda Hamiltonian near the origin, when expressed in these two coordinate systems, coincide up to order $4$. Hence Proposition \ref{localtheorem} provides us with the Taylor series expansion of $\hcalba(I) = \frac{N\b^2}{2} + H_\a(I)$ in terms of the actions
\begin{displaymath}
  I = (I_k)_{\1N1}, \qquad I_k = \frac{x_k^2 + y_k^2}{2}
\end{displaymath}
up to order $2$. In view of (\ref{g0def}), (\ref{g2lambdadef}), and (\ref{ping4ping3f3}) one has
\begin{displaymath}
  H_\a(I) = N \a^2 + \a \sum_{k=1}^{N-1} s_k I_k + \frac{1}{4N} \sum_{k=1}^{N-1} I_k^2 + O(I^3).
\end{displaymath}
This proves Theorem \ref{bnftodatheorem}.
\end{proof}

\appendix

\end{document}